\documentclass[12pt]{article}
\usepackage{graphicx}
\usepackage{color}
\usepackage{amssymb}
\usepackage{amsmath}
\usepackage{mathabx}
\usepackage{upgreek}
\usepackage{bbm}
\usepackage{multirow}
\usepackage{multicol}
\usepackage{amscd}
\usepackage{ipa}
\usepackage{mfpic}
\usepackage{verbatim}
\usepackage{cancel}
\usepackage{chemarrow}
\usepackage[linktocpage=true,colorlinks = true,
            linkcolor = darkviolet,
            urlcolor  = blue,
            citecolor = blue,
            anchorcolor = magenta]{hyperref}
\usepackage{setspace}

\definecolor{cream}{rgb}{.97, .95, .88}
\definecolor{darkcream}{rgb}{1., .88, .5}
\definecolor{lightpink}{rgb}{0.98, 0.88, 0.87}
\definecolor{lightwhite}{rgb}{1., 0.98, 0.95}
\definecolor{lightsalmon}{rgb}{1., 0.95, 0.90}
\definecolor{lightviolet}{rgb}{0.9, 0.8, 0.9}
\definecolor{lightgray}{rgb}{.96, .96, .96}  
\definecolor{lgray}{rgb}{.75, .75, .75}
\definecolor{LemonChiffon}{rgb}{0.95, 1., 0.7}
\definecolor{lightolivegreen}{rgb}{0.84, 0.89, 0.25}
\definecolor{lightgreen}{rgb}{.664, 1., .52}
\definecolor{llgreen}{rgb}{.900, .983, .960}
\definecolor{tristle}{rgb}{0.87, 0.67, 0.77} 
\definecolor{pink}{rgb}{0.95, 0.45, 0.75}
\definecolor{magenta}{rgb}{1., 0, 1.}
\definecolor{violet}{rgb}{0.9, 0.20, 0.85}
\definecolor{darkolivegreen}{rgb}{0.55, 0.65, 0.35}
\definecolor{maroon}{rgb}{0.7, 0.26, 0.56}
\definecolor{lightmaroon}{rgb}{0.85, 0.38, 0.58}
\definecolor{darkmaroon}{rgb}{0.604, 0.169, 0.451}
\definecolor{ddarkmaroon}{rgb}{0.2, 0.03125, 0.150}
\definecolor{mediumorchid}{rgb}{0.8, 0.33, 0.83}
\definecolor{mediumorchidd}{rgb}{1., 0.33, 0.63}
\definecolor{darkgreen}{rgb}{0.1, 0.6, 0.13}
\definecolor{lightyellow}{rgb}{1., 1., 0.82}
\definecolor{turquoise}{rgb}{0.042, 0.586, 0.512}
\definecolor{turquoisel}{rgb}{0.66, 0.94, 0.83}
\definecolor{darkturquoise}{rgb}{0.21, 0.55, 0.50}
\definecolor{coral}{rgb}{1., 0.6, 0.21}
\definecolor{lightorange}{rgb}{1., 0.88, 0.75}
\definecolor{orangered}{rgb}{1., 0.5, 0.}
\definecolor{orange}{rgb}{1., 0.65, 0.1}
\definecolor{orangel}{rgb}{1., .85, .3}
\definecolor{darkorange}{rgb}{0.875, 0.4, 0.204}
\definecolor{ddarkorange}{rgb}{.675, .218, .05}
\definecolor{bluesky}{rgb}{0.48, 0.53, 1.}
\definecolor{gold}{rgb}{1., 0.85, 0.25}
\definecolor{goldd}{rgb}{0.95, 0.75, 0.05}
\definecolor{darkviolet}{rgb}{0.54, 0.04, 0.84}
\definecolor{ddarkviolet}{rgb}{.382, .063, .657}
\definecolor{lightblue}{rgb}{0.30, 0.86, 0.89}
\definecolor{LightBlue}{rgb}{0.68, 0.85, 0.9}
\definecolor{lblue}{rgb}{0.78, 0.90, 0.95}
\definecolor{darkblue}{rgb}{.105, .308, .707}
\definecolor{lightmaroon}{rgb}{0.85, 0.38, 0.58}
\definecolor{darkmaroon}{rgb}{0.604, 0.169, 0.451}
\definecolor{darkpink}{rgb}{0.879, 0.020, 0.766}
\definecolor{ddarkpink}{rgb}{0.738, 0.195, 0.406}
\definecolor{grey}{rgb}{0.717, 0.717, 0.717}
\definecolor{lightgrey}{rgb}{0.800, 0.800, 0.800}
\definecolor{brown}{rgb}{0.740, 0.323, 0.182}
\definecolor{redbrown}{rgb}{.575, .158, .05}
\definecolor{darkbrown}{rgb}{0.34, 0.25, 0.05}
\definecolor{orangebrown}{rgb}{0.433, 0.262, 0.06}
\definecolor{pinkl}{rgb}{1., 0.788, 0.918}
\definecolor{salmon}{rgb}{1., 0.66, 0.5}
\definecolor{lightbrown}{rgb}{0.703, 0.508, 0.121}

\def\Journal#1#2#3#4{{#1} {\bf #2}, (#3) #4}

\def\Name#1#2 {{ #1 }{#2}}

\def\CMP{\em Commun. Math. Phys.}

\def\JPC{\em J. Phys. Conf. Ser.}

\def\JPU{\em J. Phys. (USSR)}

\def\MDU{\em MDPI Universe J.}

\def\PLA{{\em Phys. Lett.}~\emph{A}}

\def\PRD{{\em Phys.~Rev.}~\emph{D}}

\def\PRL{\em Phys. Rev. Lett.}

\def\PRX{{\em Phys.~Rev.}~\emph{X}}

\def\SYM{\em Symmetry}

\def\be{\begin{equation}}
\def\ee{\end{equation}}
\def\bea{\begin{eqnarray}}
\def\eea{\end{eqnarray}}
\def\bes{\begin{equation*}}
\def\ees{\end{equation*}}
\def\beas{\begin{eqnarray*}}
\def\eeas{\end{eqnarray*}}

\def\tr{\text{tr}}

\def\bm{\mathcal B}

\def\gm{\mathcal G}
\def\hm{\mathcal H}
\def\lm{\mathcal L}
\def\mm{\mathcal M}
\def\nm{\mathcal N}

\def\hA{\hat{A}}

\def\hH{\hat{H}}

\def\hL{\hat{L}}
\def\hO{\hat{O}}

\def\hrho{\hat{\rho}}
\def\sD{\cancel{D}}
\def\sqgr{$SU(\infty)$-QGR}
\def\suinf{{SU(\infty)}}
\def\suinfa{{su(\infty)}}

\begin{document}
\begin{center}
{\large \bf $\mathbf {\suinf}$-QGR Quantumania: Everything, Everywhere, All At Once} \\
\bigskip
Essay written for the Gravity Research Foundation, 2023 Awards for Essays on Gravitation


Houri Ziaeepour$^{a,b}$ \\
{\tt houriziaeepour@gmail.com or hz@mssl.ucl.ac.uk} \\
\end{center}

{\small a) Institut UTINAM, CNRS UMR 6213, Observatoire de Besan\c{c}on, Universit\'e de Franche Compt\'e, 41 bis ave. de l'Observatoire, BP 1615, 25010 Besan\c{c}on, France, \\
b) Mullard Space Science Laboratory, University College London, Holmbury St. Mary, GU5 6NT, Dorking, UK} \\


\begin{abstract}
\sqgr~is a quantum approach to Universe and gravity. Its main assumption is infinite mutually 
commuting observables in the Universe, leading to representation of $\suinf$ by its Hilbert 
spaces and those of its subsystems. The Universe as a whole is static, topological, and 
characterized by two continuous parameters. Nonetheless, quantum fluctuations induce 
clustering and finite rank internal symmetries, which approximately divide the Universe to 
infinite interacting subsystems. Their Hilbert space depends on an additional dimensionful 
parameter, and selection of a subsystem as clock induces a relative dynamics, with $\suinf$ 
sector as gravity. The Lagrangian defined on the (3+1)-dimensional parameter space is 
Yang-Mills for both symmetries. When quantumness of gravity is undetectable, it is perceived 
as curvature of an effective spacetime.
\end{abstract}

\thispagestyle{empty}
\pagebreak
\setcounter{page}{1}
\paragraph*{\bf Introduction:} 
Despite tremendous success of general relativity and Einstein theory of gravity, several 
fundamental questions about the nature of spacetime and gravitational interaction remain 
unanswered or ambiguous. Among them are:
\begin{description}
\item {\bf \it What is the nature of spacetime ?} 
Failure of attempts to observe physical signature of an empty space - the ether - was the main 
motivation behind the development of special and ultimately general relativity. According to 
Einstein in absence of gravitational force, that is the nontrivial spin-2 graviton field 
interpreted as a metric, the empty spacetime does not have a physical 
existence~\cite{einsteinether}. In other words, the empty space does not exist. 
\item {\bf \it Why is gravity related to spacetime ?}
If Einstein opinion is correct, why and how do gravitons introduce the spacetime and 
characterize its geometry ? Despite these ambiguities, modern approaches to Quantum GRavity 
(QGR), such as string theory and Loop QGR (LQG), treat the space(time) as an independent 
physical entity. In string and matrix theories spacetime consists of non-compactified fields 
having special configurations, such as a D-brane, in a higher dimensional space. LQG 
concentrates on the space (without time) as the main physical entity for gravitation and 
tries to quantize it independently from its matter content, which is the source of gravity.
\item {\bf \it What does determine the dimension of spacetime ?} 
Why do we perceive the Universe in 3 dimension plus time ? In general relativity and in many 
quantum gravity proposals the dimension of spacetime is taken for granted and no explanation 
is sought.
\item {\bf \it Why do the representation of spacetime and internal symmetries by elementary 
particles are so different ?} 
Non-gravitational fundamental forces are described by Yang-Mills gauge theories. Observations 
show that density matrices of matter fields and force mediators are in the same representation 
of internal symmetries. There is not an analogous relation between representations of 
spacetime related Lorentz symmetry by matter fields and by gravitons. Why ? In other words, 
why is the mediator of gravity a spin-2 field which makes its canonical quantization 
nonrenormalizable, rather than spin-1 like other interactions ?
\end{description}
In this essay we outline the structure and properties of a foundationally quantum model 
- dubbed \sqgr~- for the Universe and its contents, in which a universal force similar to 
gravity emerges, and above questions are answered. In addition, when quantumness of gravity 
is not observable, the model is fully consistent with Einstein gravity. \sqgr~has been 
first reported in~\cite{houriqmsymmgr}, compared with some of popular QGR models 
in~\cite{hourisqgrcomp}, and more thoroughly investigated in~\cite{hourisqgr}.

\paragraph*{Axioms and Hilbert space of the Universe: }
We begin by designing a quantum Universe based on three well motivated assumptions:
\setcounter{enumi}{0}
\renewcommand{\theenumi}{\Roman{enumi}}
\begin{enumerate}
\item Quantum mechanics is valid at all scales and applies to every entity, including the 
Universe as a whole;\label{uniaxiom1}
\item Every quantum system is described by its symmetries and its Hilbert space represents 
them;\label{uniaxiom2}
\item The Universe has an infinite number of independent degrees of freedom, which are 
associated to mutually commuting quantum observables.\label{uniaxiom3} 
\end{enumerate}
Although these axioms may seem trivial, in this essay we show that they are sufficient 
for describing a universe with a universal quantum interaction between its subsystems - 
contents. We interpret it as gravity. 

In a quantum system with above properties, both the Hilbert space $\hm_U$ and space of 
(bounded) linear operators $\bm[\hm_U]$ are infinite dimensional and represent $\suinf$ 
group. It is known that representations of $\suinf$ are homomorphic to area preserving 
diffeomorphism of 2D compact Riemann surfaces~\cite{suninfhoppthesis}, and are classified by 
their genus. They present different realizations of this quantum Universe, and in each case 
coordinates $(\theta, \phi)$ of the 2D Riemann surface, that we call the {\it diffeo-surface}, 
parameterize states of $\hm_U$ and operators belonging to $\bm[\hm_U]$. In particular, 
generators of the $\suinfa$ Lie algebra realized by $\bm[\hm_U]$ have a description as a 
functional of spherical harmonic functions: 
\bea
&& \{\hL_{lm}(\theta, \phi),~\hL_{l'm'}(\theta, \phi)\} = -i \frac{\hbar}{cM_P} 
f ^{l"m"}_{lm,l'm'} \hL_{l"m"}(\theta, \phi) \label{lapp}  \\
&& \hL_{lm} (\theta, \phi) = i\hbar ~ \sqrt{|g^{(2)}|} \epsilon^{\mu\nu} 
(\partial_\mu Y_{lm}(\theta, \phi)) \partial_\nu, \quad \quad {\mu,\nu \in \{\theta,\phi \}}, 
\quad l\geqslant 0,~|m| \leqslant l \label{lharminicexp}
\eea
Moreover, spherical harmonics $Y_{lm}$'s satisfy Poisson algebra, which is homomorphic to 
$\suinfa$, and structure constants $f^{l"m"}_{lm,l'm'}$ are related to 3j 
symbols~\cite{suninfhoppthesis}.

In \sqgr~the non-Abelian algebra (\ref{lapp}) replaces the usual quantization 
relations~\cite{qmmathbook,qgrnoncommut}. The normalization chosen for operators $\hL_{lm}$ 
in (\ref{lapp}) is such that if the Planck constant $\hbar\rightarrow 0$ or the Planck mass 
$M_P\rightarrow \infty$, the algebra becomes Abelian and homomorphic to 
$\bigotimes^{N \rightarrow \infty} U(1)$ of classical observables. This means that it cannot 
used for quantization of the model. Therefore, in \sqgr~quantumness and gravity are 
inseparable.

\paragraph*{Contents of the Universe: }
Despite infinite number of observables in this quantum Universe, it is not straightforward 
to distinguish subsystems - the Universe's contents. Indeed, a divisible quantum system must 
fulfill specific conditions~\cite{sysdiv}. In particular, linear operators applied to its 
state should consist of mutually commuting subsets $\{\hA_i\}$'s, where each 
subset represents an {\it internal} symmetry $G_i$. Of course, the symmetry of subsets do not 
need to be different. Subsystems representing the same symmetry will be indistinguishable 
and their number will present their multiplicity. 

To understand how clustering and an approximately division of the Universe to commuting 
subspaces may arise, consider the Universe in a pure completely coherent state in an arbitrary 
basis, that is $\hrho_U = \hrho^{CC} \equiv \nm \sum_{a,b} |a\rangle \langle b|$. This state has 
maximum symmetry, in the sense that in the basis $\{|a\rangle\}$ all eigen states have the 
same probability of occurrence. Random acting of operators $\hO \in \bm[\hm_U]$ on the 
$\hrho_U$ - in other words quantum fluctuations - reduce the symmetry of $\hrho_U$ by 
increasing probability of some eigen states and decreasing those of others. Repetition of 
such operations is more likely to lead to clustering and emergence of approximately orthogonal 
blocks in $\hrho_U$, rather than restoring the unique $\hrho^{CC}$ state. Although, due to the 
assumed $\suinf$ symmetry and homomorphism $\bm[\hm_U] \cong \suinf$, transformations of 
$\hrho_U$ do not change the global dynamics of the Universe, they induce a structure and a 
concept of {\it locality} to its state. 

According to complementarity condition for division of a quantum system~\cite{sysdiv} 
$\otimes_i \{\hA_i\} \cong End (\bm[\hm_U])$. Moreover, considering properties of $\suinf$, 
in particular $\suinf^n \cong \suinf, ~\forall n$, if $G_i$'s have infinite ranks, they 
would be homomorphic to $\suinf$ and subsystems would be indistinguishable from the whole 
Universe. Therefore, $G_i$'s must have finite ranks and the number of subsystems must be 
infinite. Due to quantum superposition the $Z^{N \rightarrow \infty}$ symmetry of subsystems 
with similar internal symmetry is uplifted and their Hilbert spaces effectively represent 
$\gm_i = \suinf \times G_i$. Thus, all subsystems interact through their common $\suinf$ 
symmetry. In addition, this mutual relationship manifests itself as quantum entanglement of 
every subsystem with the rest of the Universe~\cite{hourisqgr}. We refer to this property 
as {\it global entanglement}.

\paragraph*{Emergence of spacetime: }
Area of diffeo-surface of one $\suinf$ representation is irrelevant. This means that 
area preserving diffeo-surfaces represent 
$\suinf \times \mathbb{R} \cong \suinf \times U(1) = U(\infty)$. Once there are more than one 
quantum systems representing $\suinf$, areas or sizes of their diffeo-surfaces can be compared 
with each others and become a dimensionful relative observable, determined with respect to 
a reference subsystem - an arbitrary standard ruler. Notice that comparability of the areas 
of diffeo-surfaces does not mean that their values are considered as being fixed. The area 
or size becomes rather a new relative measurable, in the same way as in quantum physics 
without gravity, where the distance from a reference is an observable and state of a quantum 
system can be in superposition of its eigen states. Consequently, in addition to 
$(\theta, \phi)$ parameters, states of Hilbert space $\hm_s$ of subsystems and operators in 
$\bm[\hm_s]$ depend on a dimensionful parameter $r$. For instance, if diffeo-surfaces are 
embedded in $\mathbb{R}^{(3)}$, the value of $r$ may be chosen to be the geometric distance 
from a reference point. Nonetheless, any other choice would be equally valid. 

Division of the Universe to subsystems makes it possible to choose one of them as a quantum 
clock and to associate a time parameter to variation of its state or more generally to those 
of its observables. Then, comparison of subsystems states and their variations with variation 
of time parameter provides a relative dynamics \`a la Page \& Wootters~\cite{qmtimepage} or 
equivalent methods~\cite{qmtimedef}. Specifically, dynamics arises in an operational manner:  
A random application of an operator $\hO$ to the state $\hrho_s$ of a subsystem - 
a quantum fluctuation - changes its state to $\hO \hrho_s \hO^\dagger$. The global 
entanglement convoys this change to other subsystems, including the clock, which in turn 
have their own change of state both coherently and due to interaction with other subsystems. 
Therefore, an arrow of time arises automatically and persists eternally. Although inverse 
processes are in principal possible, giving the infinite number of subsystems and operations, 
and their entanglement, inverting the arrow of time is extremely improbable.

Considering ensemble of parameters that emerge after division of the Universe to subsystems 
and introduction of dynamics, quantum state of a subsystem can be written as:
\be
|\psi_s \rangle = \sum_{t, r, \theta, \phi, \alpha} \psi^\alpha_s (t, r, \theta, \phi) ~ 
|t, r, \theta, \phi \rangle \otimes |\alpha \rangle  \label{subsysstate}
\ee
where $|t, r, \theta, \phi \rangle \otimes |\alpha \rangle$ is an eigen basis for the 
Hilbert space $\hm_s$, and $\alpha$ is a collective parameter for representation of the 
internal symmetry $G$ by a subsystem. If we consider state of the density matrix 
$\hrho_s(t)$ for a given value of time parameter $t$, its Hamiltonian operator $\hH_s$ is 
defined by Schr\"odinger equation $d\hrho_s (t) /dt = -i/\hbar [\hH_s, \hrho_s]$. 

Although the origin of parameters $(t, r, \theta, \phi) \in \Xi$ that characterize $\suinf$ 
symmetry and dynamics of subsystems are very different, due to arbitrariness of the Hilbert 
space basis, quantum superposition, and entanglement of clock and reference with other 
subsystems according to the global entanglement, in general their eigen states cannot be 
factorized. Hence, a transformation of the Hilbert space $\hm_s$'s basis is equivalent to a 
diffeomorphism in its parameter space $\Xi$. In addition, to any pair of continuous 
parameters $x \equiv (t, r, \theta, \phi) \in \Xi$ an algebra homomorphic to (\ref{lapp}) can 
be associated. This implies that every pair of parameters can be considered as parameters 
of a representation of $\suinf$ symmetry. Consequently, up to an unobservable global 
rescaling a deformation - diffeomorphism - of $\Xi$ can be compensated by $\suinf$ 
transformations~\cite{hourisqgr}, and geometry of the parameter space $\Xi$ is arbitrary and 
irrelevant for physical observables. 

\paragraph*{Classical geometry from quantum properties: }
The space $\Xi$ of continuous parameters of the Hilbert space of subsystems looks like 
the classical spacetime. Thus, irrelevance of its geometry might be interpreted as being 
in contradiction with observations. Here we show that this similarity is superficial and 
the perceived classical spacetime traced by observations of classical systems is not $\Xi$, 
but an effective path in this space. 

Continuity of parameters $x \in \Xi$ of the Hilbert space $\hm_{U_s}$ means that they 
have their dual in the dual Hilbert space $\hm^*_{U_s}$, and together satisfy the usual 
uncertainty relations. In particular, Mandelstam-Tamm speed limit on the variation of a 
quantum state~\cite{qmspeed,qmspeedlimit,qmspeedlimgen} applies to subsystems and their 
ensemble. Consider an infinitesimal variation of the state 
$\hrho_s \rightarrow\hrho_s + \delta\hrho_s$ after tracing out the contribution of internal 
symmetries. The Mandelstam-Tamm inequality can be written as:
\be
\Lambda \frac{2}{\hbar^2} Q(\hH, \hrho_s) dt^2 \geqslant -\frac{\Lambda}{\hbar^2} 
\tr [\sqrt{\hrho_1}, \hH]^2 dt^2 = \Lambda \tr (\sqrt{\delta\hrho_1} 
\sqrt{\delta\hrho_1}^\dagger) \equiv \Lambda ds_{WY}^2 \equiv ds^2 \label{separation}
\ee
where $ds_{WY}$ is the separation between the two states according to the Wigner-Yanase 
skew information~\cite{wigneryanaseqminfo,qmspeedlimgen}. The arbitrary dimensionful 
constant $\Lambda$ gives $ds^2 \equiv g_{\mu\nu}(x) dx^\mu dx^\nu$ the dimension of 
affine separation, and $x^\mu$ can be interpreted as an average parameter in the 4D $\Xi$
space presenting the average path of the state variation. Notice that $ds$ is independent 
of the choice of basis for the Hilbert space and parameterization of its states. The 
effective metric $g_{\mu\nu}$ in (\ref{separation}) is clearly related to the quantum state 
of subsystems and their variation. In addition, the inequality in (\ref{separation}) can be 
used to prove that the signature of $g_{\mu\nu}$ must be negative for fully separable 
$\hrho_s$ and $\hrho_s + \delta\hrho_s$~\cite{houriqmsymmgr,hourisqgr}. Thus, we conclude 
that what we perceive as a (3+1)D classical curved spacetime through its gravitational 
effects is indeed this effective geometry induced by the $\suinf$ symmetry of the Universe 
and its contents. 

If the contribution of internal symmetries in the density matrix is not traced, one can 
define a metric and an affine parameter that includes also the parameters of internal 
symmetries. However, parameters characterizing representations of finite rank Lie groups 
are usually discrete. Therefore, in contrast to some other QGR proposals, 
{\it extra-dimensions} of this extended geometry would not be continuous.

Although, relations in (\ref{separation}) are not specific to \sqgr, this model relates them 
to $\suinf$ symmetry and quantum gravity. In semi-classical gravity an analogous relation is 
established through the expectation value of a quantum energy-momentum tensor, which in 
contrast to $\hrho_s$ and its variation, is not well defined unless operator ordering is 
applied~\cite{devacuum0}.

\paragraph*{Dynamics: }
Dynamics of systems in quantum mechanics and QFT are usually inspired from classical 
limit of the models. This approach cannot be directly applied to \sqgr, because as we 
discussed earlier, if $\hbar \rightarrow 0$, the model becomes trivial. On the other 
hand, considering the fact that physical systems have tendency to approach to an 
equilibrium state, variational principle can be applied to \sqgr, if a suitable functional 
of parameters and observables of the model - a Lagrangian - can be found such that it 
respect symmetries of the Hilbert space and its parameters.

In the case of Universe as a whole the symmetry is $\suinf$ and observables can be expanded 
with respect to generators of $\suinfa$ algebra, namely $\hL_{lm}$. As trace of 
multiplications of generators are invariant under $\suinf$, in a symmetry invariant 
functional generators of $\suinf$ should be traced, both over indices $(l,m)$ and over 
continuous parameters $(\theta, \phi)$ in a reparameterization independent manner. It can 
be shown that at lowest order of traces, the Lagrangian functional must have a $\suinf$ 
Yang-Mills form:
\bea
&& \lm_U = \int d^2\Omega \sqrt{|\upeta|} \biggl [ ~\frac{1}{2} ~ 
\tr (F^{\mu\nu} F_{\mu\nu}) + \frac {1}{2} \tr (\sD \hrho_U) \biggr ], \nonumber \\
&& F_{\mu\nu} \equiv F_{\mu\nu}^a \hL^a = [D_\mu, D_\nu], \quad 
D_\mu \equiv (\partial_\mu - \Gamma_\mu) \mathbbm {1} - i A_\mu^a \hL^a \label{yminvardef}
\eea
where $a \equiv (l,m)$, $\upeta$ is determinant of the metric of diffeo-surface; and 
$\Gamma_\mu$ is the corresponding connection. This Lagrangian is static, and it can be shown 
that its ground state is trivial~\cite{houriqmsymmgr}. Moreover, as we discussed earlier, 
variation of the metric can be compensated by a $\suinf$ gauge transformation. Thus, 
(\ref{yminvardef}) is topological. In particular, as there is only one topological class 
in 2D, namely the Euler characteristic 
$\chi (\mm) \equiv 2 - \gm (\mm) = 1/4\pi \int d^2\Omega ~ \mathcal{R}^{(2)}$, 
the gauge term $\tr (F^{\mu\nu} F_{\mu\nu}) ~ \propto ~ \mathcal{R}^{(2)}$. This relation shows 
how the scalar curvature of the parameter space - the diffeo-surface -  emerges in \sqgr. As 
$\hrho_s \in \bm[\hm_U]$, it can be also expanded with respect to $\hL_{lm}$'s and similar 
arguments about its independence of the metric of parameter space applies.

Similar arguments as above also apply to subsystems of the Universe, which in addition to 
$\suinf$ are invariant under a finite rank group $G$. They lead to a Lagrangian for 
subsystems which is Yang-Mills in both $\suinf$ and $G$ over the (3+1)D parameter space 
$\Xi$, related to dynamics and $\suinf$ symmetry of subsystems:
\be
\lm_{U_s} = \int d^4x \sqrt{|\upeta|} ~ \biggl [\frac{1}{16\pi L_P^2} 
\tr (F^{\mu\nu} F_{\mu\nu}) + \frac{\lambda}{4} \tr (G^{\mu\nu} G_{\mu\nu}) + 
\frac {1}{2} \sum_s \tr (\sD \hrho_s) \biggr ] \label{yminvarsub}
\ee
It is crucial to remind that Yang-Mills models are renormalizable. Therefore, \sqgr~ 
satisfies this crucial criteria, which is an insurmountable problem for many QGR proposals.

\paragraph*{Classical limit of gravity: }
As explained earlier, the usual definition of classical limit as $\hbar \rightarrow 0$ 
makes \sqgr~trivial. Therefore, we define classical limit as the case where quantum effects 
of gravity, in particular the exact form of the pure $\suinf$ gauge term in 
(\ref{yminvarsub}) is not discernible for the observer and is perceived as 
a scalar function of continuous parameters $x \in \Xi$. According to a theorem by 
A. L. Besse~\cite{curvaturfunc} in 3 or higher dimension spaces any scalar function is 
proportional to scalar curvature for a specific definition of the metric. Therefore, in 
the classical limit, the first term in (\ref{yminvarsub}) is interpreted as scalar 
curvature of the spacetime, which its metric is determined by solving dynamic equation 
obtained from applying variational principle to $\lm_{U_s}$ with the first term considered 
to be the scalar curvature. Thus, $\lm_{U_s}$ takes the form of a Yang-Mills QFT for $G$ 
in a classical background spacetime ruled by Einstein-Hilbert action, which so far is 
consistent with all observations. Indeed, due to smallness of the Planck length $L_P$, 
higher order quantum corrections in the effective action are strongly suppressed.

The Lagrangian $\lm_{U_s}$ does not explicitly include a cosmological constant. Nonetheless, 
several processes specific to \sqgr~can behave as an effective cosmological constant or 
more generally dark energy. They are discussed in~\cite{hourisqgr}.

\paragraph*{Observable signatures of \sqgr: } The most discriminative properties of this 
model are vector nature of gravity mediator boson at quantum level, Yang-Mills form of its 
interaction with matter, and $\suinf$ as its gauge symmetry. Although some QGR 
models, such as M-theory and AdS/CFT are based on the duality between gravity and a gauge 
QFT, the three distinctive properties of \sqgr~are not simultaneously present. 

As a non-Abelian Yang-Mills theory, one might add a dual gauge term to Lagrangian 
$\lm_{U_s}$, and breaks the parity symmetry. There is no natural {\it axion} in \sqgr~to 
restore symmetry. Therefore, observation of parity violation in gravitational sector might 
encourage a model similar to \sqgr. However, according to the Besse theorem, such term, and 
indeed any addition to $\lm_{U_s}$ would not be in general expressible as a function of 
the same scalar curvature and effective metric. These properties may be easier to test 
than quantum processes of gravity, which at present and in near future are not accessible 
to experiments~\cite{qgrtestgw}.However, they would not be indisputable evidence of \sqgr.

\paragraph*{Quantumania: Everything, Everywhere, All at Once: }
The \sqgr~Quantumania is static and topological - Everything happens Everywhere and All at 
Once - as there is no {\it once}. But, {\it once} and a 3D {\it where} emerge when 
randomly {\it things} are formed, and can be distinguished - not by a supreme being out of 
Quantumania or inside it, but by infinite small and big things emerging randomly: 
{\it In Quantumania reciprocity is the rule}. Everything is everywhere; no division, 
no privacy, and no place to hide. Everyone is entangled with the rest. Time is relative - 
an order of changes - thus, in principle reversible. But, how to invert the state of 
infinite entangled things, specially when uncertainty and non-commutativity reign ?! Thus, 
the arrow of time is eternal. And the Multiverse ? What multiverse ?! Connect them and be 
in a new one. But how to know that it is new ?! Sum of infinities is again infinity ! 

\pagebreak

\end{document}